\documentclass[11pt]{article}

\usepackage[final]{acl}

\usepackage{times}
\usepackage{latexsym}
 \usepackage{amsmath}
\usepackage[T1]{fontenc}

\usepackage[utf8]{inputenc}

\usepackage{microtype}

\usepackage{inconsolata}

\usepackage{graphicx}
\usepackage{booktabs}
\usepackage{placeins}
\usepackage{dblfloatfix}
\usepackage{float}
\usepackage{needspace}

\usepackage{xcolor}

\title{Tracing a Sparse Emotion-Control Circuit in LLM-Based Text-to-Speech}

\author{
  \textbf{Hongfei Du},
  \textbf{Jiacheng Shi},
  \textbf{Yanfu Zhang},
  \textbf{Ye Gao}
  \\
  {\normalfont Department of Computer Science, William \& Mary, Williamsburg, USA}
  \\
  {\normalfont\texttt{\{hdu02, jshi12, yzhang105, ygao18\}@wm.edu}}
}

\begin{document}
\maketitle

\begin{abstract}
LLM-based text-to-speech (TTS) models can generate emotionally expressive
speech, but how reference emotion is routed through the model and realized in
decoded speech remains unclear. We introduce two emotion-sensitive metrics for
matched neutral and emotional syntheses---a codec trajectory score and a late
residual direction score---and use them to score activation-patching
interventions. Under controlled matched-reference conditions, this analysis
identifies a sparse source-to-readout component-level circuit: 23--27 attention
heads and MLPs per emotion, roughly 5\% of the components considered, recover or
suppress 74--88\% of the late emotion-readout shift on held-out cases. The
circuit combines a shared component backbone with emotion-specific components;
cross-emotion activation swaps reduce the target readout in 47 of 48 cases. In
decoded speech, the same intervention produces consistent changes in pitch,
energy, and spectral brightness over 24 matched pairs per emotion. A
readout-matched residual-direction baseline produces only 17--27\% of the
intervention's pitch effect, showing that internal readout movement alone does
not explain the decoded acoustic changes. These results trace a compact causal
route from reference-derived prefix information to emotion-relevant properties
of generated speech.
\end{abstract}

\section{Introduction}
LLM-based text-to-speech (TTS) models have recently shown strong capability for
generating natural and emotionally expressive speech from text, speaker
information, and reference prompts
\citep{wang2023neural,du2024cosyvoice,wang2025sparktts,deng2025indextts,zhou2025indextts2}.
Controllable style and emotion generation is especially important for spoken
dialogue agents, virtual humans, and health-support interfaces
\citep{picard1997affective,bickmore2005establishing}, because vocal emotion and
prosody shape how listeners perceive intent, engagement, and empathy
\citep{scherer2003vocal,juslin2003communication}.
Recent TTS systems support such control through reference prompts, style
descriptions, or explicit emotion conditioning
\citep{guo2022prompttts,yang2023instructtts,zhou2025indextts2,yang2025emovoice}.
This capability raises a mechanistic question: how does reference emotion enter
an LLM-based TTS model, and which internal components carry its effect toward
generated speech?

Prior mechanistic studies of language models often study behaviors with
localized output positions
\citep{olah2020zoom,elhage2021mathematical}. In factual recall, causal tracing,
and automated circuit discovery, interventions can often be scored at answer
tokens or other local prediction sites
\citep{vig2020causal,meng2022locating,heimersheim2024activation,conmy2023automated}.
This scoring strategy does not transfer directly to LLM-based TTS. Reference
emotion affects a long sequence of generated speech tokens, which are then
transformed by acoustic decoding before they become audible. A decoded-speech
emotion score can verify that decoded speech moves toward a target emotion, but it
does not reveal which hidden states, components, or generation steps caused the
change. Mechanistic analysis of TTS therefore needs measurements that link
internal activations to codec-level changes and decoded-speech outcomes.

Our causal tracing workflow has four steps. We first build matched neutral and
emotional synthesis pairs, then localize where reference emotion enters the
conditioning prefix, trace its effect through attention heads and MLPs to late
emotion readouts, and finally test whether the selected components causally
change decoded speech. Together, these steps turn a decoded-speech emotion change
into a source-to-readout causal localization problem.

\begin{figure*}[t]
  \centering
  \includegraphics[width=\linewidth]{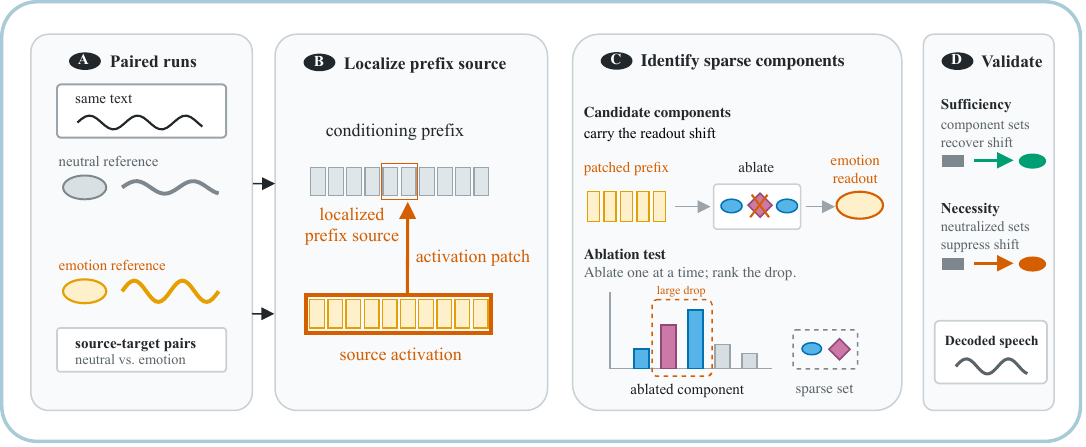}
\caption{\textbf{Causal tracing localizes a sparse source-to-readout component-level emotion circuit within a reference-conditioned LLM-based TTS model.}
  Matched neutral/emotional runs localize a conditioning-prefix source, trace its effect through
  sparse attention-head and MLP sets to late emotion readouts, and validate the
  pathway with causal and decoded-speech tests.}
  \label{fig:causal_tracing_overview}
\end{figure*}

To causally localize the reference emotion effect, we construct matched neutral
and emotional syntheses that keep text and speaker setting fixed while changing
the emotion reference. For these pairs, we design and validate two
emotion-sensitive metrics: a codec trajectory score over generated token
trajectories and a late residual direction score over generation states. These
metrics let us score activation-patching interventions that replace selected
internal activations from one synthesis trajectory with matched activations from
another. These scores support prefix localization, component-level tracing,
matched controls, component specificity checks, and decoded-audio validation.
Figure~\ref{fig:causal_tracing_overview} summarizes this complete workflow.

Our activation-patching analysis identifies a sparse source-to-readout
component-level circuit under controlled matched-reference conditions. The
circuit starts at a shared conditioning-prefix source and passes through 23--27
selected attention heads and MLPs per emotion, roughly 5\% of the components
considered, to late emotion readouts. On held-out cases, these components recover
or suppress 74--88\% of the readout shift and exceed matched random sets in every
emotion and intervention direction. The circuit has a shared computational
backbone together with emotion-specific components: shared-only sets recover
79--86\% of the full-set effect, while cross-emotion donor activations reduce the
target readout in 47 of 48 cases. Finally, paired analyses over 24 decoded
waveforms per emotion show consistent changes in pitch, energy, and spectral
brightness. A residual-direction baseline matched to the internal readout shift
produces only 17--27\% of the selected components' pitch effect, connecting the
localized components to acoustic changes beyond a global emotion direction.

Our main contributions are:
\begin{itemize}
    \item We design and validate two emotion-sensitive metrics for causal
    localization in TTS: a codec trajectory score and a late residual direction
    score. Unlike prior circuit work that scores answer-token interventions,
    these metrics score patching across codec-generation trajectories and late
    residual states, enabling localization that can be tested against decoded
    waveform evidence.

    \item We identify a sparse source-to-readout component-level circuit for
    reference emotion. It begins at a shared conditioning-prefix source and
    passes through 23--27 selected attention heads and MLPs per emotion to late
    readouts for anger, happiness, and sadness.

    \item We show that the circuit is causal and structured. Sufficiency and
    necessity interventions recover or suppress 74--88\% of the readout shift;
    shared/specific decomposition and cross-emotion swaps reveal a shared
    backbone carrying emotion-dependent activation content.

    \item We connect the internal circuit to decoded speech. The selected
    components produce consistent emotion-relevant acoustic changes and
    substantially outperform readout-matched residual-direction steering at
    transferring the internal shift to waveform pitch.
\end{itemize}

\section{Related Work}
\label{sec:related_work}

\paragraph{Emotion control in TTS.}
Emotion-controllable TTS aims to synthesize speech whose affect can be specified or guided by the
user. Existing systems expose this control through natural-language style descriptions, explicit
emotion labels, intensity controls, reference prompts, or reinforcement-learning objectives
\citep{guo2022prompttts,yang2023instructtts,zhang2025proemo,yang2025emovoice,li2026emorltts}. Modern
TTS systems often synthesize speech by generating discrete semantic or acoustic tokens before waveform
decoding \citep{borsos2023audiolm,wang2023neural}. Recent LLM-based TTS systems, including CosyVoice,
Spark-TTS, IndexTTS, and IndexTTS2, build on this token-generation design and support controllable
generation through attribute prompts, style references, or emotion inputs
\citep{du2024cosyvoice,wang2025sparktts,deng2025indextts,zhou2025indextts2}.
These systems demonstrate practical emotion control in generated speech, but they do not identify how
reference-conditioned emotion is represented or routed inside the TTS model. We instead ask where
reference-conditioned emotion enters LLM-based TTS generation and which internal components carry its
effect to emotion-sensitive readouts.

\paragraph{Causal circuit localization.}
Mechanistic interpretability studies often explain a model behavior by
identifying the internal states or components that causally support a chosen
metric \citep{olah2020zoom,elhage2021mathematical,olsson2022induction}.
Causal mediation, causal tracing, and activation patching test this relation by
replacing internal states under a source-target contrast
\citep{vig2020causal,meng2022locating,heimersheim2024activation}. Circuit
studies then refine this localization from residual states to attention heads,
MLPs, or paths
\citep{wang2023interpretability,conmy2023automated,goldowskydill2023path,kramar2024atp}.
Most circuit-level evidence so far comes from localized prediction tasks, such as
factual recall or indirect object identification. We study
reference-conditioned emotion in LLM-based TTS under this causal localization
perspective. Unlike localized prediction tasks, the emotion effect is distributed across a
long sequence of generated speech tokens and only becomes audible after waveform
decoding. We therefore need intervention metrics that score how changes in
prefix and generation states affect the generated token trajectory and decoded
speech.

\paragraph{Internal analyses of speech models.}
Recent work has begun to analyze internal representations in speech models.
Mechanistic studies of automatic speech recognition examine how speech
representations support transcription-related behavior
\citep{glazer2026beyond}. \citet{paek2025learning} analyze audio latent spaces
with sparse autoencoders and learn features aligned with acoustic properties
such as pitch, amplitude, and timbre. In speech synthesis, hidden-state steering
methods show that adding learned directions can change the emotion of synthesized
speech without updating model weights \citep{xie2025emosteer}. These studies
show that speech model internals can be probed or manipulated, but they do not
trace how reference-conditioned emotion flows through an LLM-based TTS model. We
instead localize a conditioning-prefix source and identify attention heads and MLPs
that carry its effect to late emotion readouts.

\section{Preliminaries}
\label{sec:preliminaries}
\paragraph{Generated speech token trajectories.}
We study the autoregressive stage that produces generated speech tokens before
waveform decoding. In LLM-based TTS, an autoregressive language-model component
predicts discrete speech tokens from text, speaker information, and an emotion
reference, and a downstream decoder converts the predicted token sequence into a
waveform
\citep{wang2023neural,du2024cosyvoice,wang2025sparktts,deng2025indextts}.
Suppressing speaker notation for clarity, we write this process as
{
\[
  p_\theta(z_t \mid x,r,z_{<t}), \qquad y = D(z_{1:T}),
\]
}
where \(x\) is the input text, \(r\) is the emotion reference, \(z_{1:T}\) is the
generated speech token trajectory, and \(D\) is the audio decoder. We intervene
during this autoregressive stage, before waveform decoding.

\paragraph{Matched synthesis pairs.}
We next define the neutral--emotion contrast that supplies source and target
trajectories. Each contrast uses matched neutral and emotional syntheses that
share the same text and speaker setting but differ in the emotion reference. The
neutral trajectory uses a neutral reference \(r^{(n)}\), while the emotional
trajectory uses a target-emotion reference \(r^{(e)}\):
{
\[
  \begin{aligned}
  \tau^{(n)} &= (x,r^{(n)},z^{(n)}_{1:T},y^{(n)}),\\
  \tau^{(e)} &= (x,r^{(e)},z^{(e)}_{1:T},y^{(e)}).
  \end{aligned}
\]
}
Here \(z_{1:T}\) is the generated speech token trajectory and \(y\) is the decoded
waveform for the same text \(x\). Because \(x\) and the speaker setting are fixed
within each pair, differences between \(\tau^{(n)}\) and \(\tau^{(e)}\) define
the reference emotion contrast while controlling for lexical content and speaker
identity. We use \(\tau^{\mathrm{src}}\) for the trajectory that supplies an
activation and \(\tau^{\mathrm{tgt}}\) for the trajectory being modified.

\paragraph{Activation patching.}
Given a source and target trajectory, activation patching tests whether copying
one internal value from the source trajectory into the target trajectory
moves the target toward the source condition. For a component site
\(q=(\ell,t,c)\), let \(a_q(\tau)\) denote the activation of component \(c\) at
layer \(\ell\) and position \(t\) on trajectory \(\tau\). A patch replaces the target activation
with its source counterpart,
{
\[
  a_q(\tau^{\mathrm{tgt}}) \leftarrow a_q(\tau^{\mathrm{src}}),
\]
}
and we score the resulting patched trajectory
\citep{vig2020causal,meng2022locating,heimersheim2024activation}. The next
section defines the scores used for this measurement.

\paragraph{Patchable states.}
Finally, we specify the internal values that can serve as patch sites in the
autoregressive model. A patch can replace a residual state, an attention-head
output, or an MLP output. At sequence position \(t\), let \(h_{\ell,t}\) be the
residual stream entering layer \(\ell\). A Transformer block updates this state
in two steps,
{
\[
  \tilde h_{\ell,t}=h_{\ell,t}+u_{\ell,t}, \qquad
  h_{\ell+1,t}=\tilde h_{\ell,t}+m_{\ell,t},
\]
}
where \(u_{\ell,t}\) is the attention update and \(m_{\ell,t}\) is the MLP update. For head-level
patches, we decompose the attention update before the output projection,
{
\[
  u_{\ell,t}=W^{O}_{\ell}\,[o_{\ell,1,t}\Vert \cdots \Vert o_{\ell,H,t}],
\]
}
where \(W^{O}_{\ell}\) is the layer-\(\ell\) attention output projection,
\(\Vert\) denotes concatenation, \(H\) is the number of attention heads, and
\(o_{\ell,j,t}\) is the output of head \(j\). Thus \(c\) can denote a residual
stream, head output, MLP output, or block state. Prefix patches use prefix
positions; generation-time component patches use generated speech-token
positions.

\section{Measuring and Localizing Causal Emotion Effects}
To localize causal emotion effects, activation-patching interventions must be
scored on matched neutral and emotional trajectories. We therefore define two
complementary metrics for this setting: a codec trajectory score and a late
residual direction score. Together, they score patched trajectories and trace
reference emotion within the TTS computation; Table~\ref{tab:emotion_metric_stability}
reports their stability.
  
\subsection{Metrics for emotion effects}
  \label{sec:two_emotion_metrics}
  Before localizing an emotion effect, we need a scalar score for how much a
  patched trajectory moves along the neutral-to-emotion contrast. For each matched
  synthesis pair \(i\), let \(\tau_i^{(n)}\) and \(\tau_i^{(e)}\) denote the
  neutral and emotional trajectories defined above, with \(\tau_i^{(e)}\)
  corresponding to the currently scored target emotion \(k\). The codec trajectory
  score represents a trajectory by the mean embedding of its predicted codec tokens,
  \(\mu(\tau)\). For target emotion \(k\), let
  \(\Delta_k^\mu=\sum_i(\mu(\tau_i^{(e)})-\mu(\tau_i^{(n)}))\). For any trajectory
  \(\tau\) paired with the neutral member of pair \(i\), such as a patched
  trajectory \(\tau_i^{\mathrm{patch}}\), we define
  {
  \[
  \begin{aligned}
    \hat d_k^\mu
    &= \Delta_k^\mu / \|\Delta_k^\mu\|_2,\\
    C_k(\tau)
    &=
    \left\langle
      \mu(\tau)-\mu(\tau_i^{(n)}),
      \hat d_k^\mu
    \right\rangle .
  \end{aligned}
  \]
  }
  Here \(C_k\) is the codec trajectory score, evaluated relative to the neutral
  member of the same matched pair. The late residual direction score
  uses \(\bar h_L(\tau)\), the mean final-layer residual state over generated
  token positions, where \(L\) denotes the final layer index. We define
  {
  \[
  \begin{aligned}
    \Delta_k^h
    &= \sum_i(\bar h_L(\tau_i^{(e)})-\bar h_L(\tau_i^{(n)})),\\
    \hat d_k^h
    &= \Delta_k^h / \|\Delta_k^h\|_2,\\
    R_k(\tau)
    &=
    \left\langle
      \bar h_L(\tau)-\bar h_L(\tau_i^{(n)}),
      \hat d_k^h
    \right\rangle .
  \end{aligned}
  \]
  }
  We use \(R_k(\tau)\) as the internal score for target emotion \(k\), with the
  same pair-relative convention. When the matched pair is clear, we suppress the
  pair index and write \(C_k(\tau)\) or \(R_k(\tau)\).

\begin{table}[t]
  \small
\centering
\begin{tabular}{lccc}
\hline
\textbf{Emotion} & \textbf{Codec pair} & \textbf{Codec split} & \textbf{Resid. cent.} \\
\hline
Anger & 0.941 & 0.998 & 0.926 \\
Happiness & 0.764 & 0.988 & 0.903 \\
Sadness & 0.903 & 0.996 & 0.874 \\
\hline
\end{tabular}
\caption{\textbf{Stability of the two emotion metrics over 50 matched pairs.}
For the codec trajectory score, Codec pair reports the mean pairwise cosine among
paired codec-token embedding shifts, and Codec split reports split-half agreement
of the global codec direction. Resid. cent. reports the mean cosine to the
final-layer residual centroid used by the late residual direction score.}
\label{tab:emotion_metric_stability}
\end{table}

\begin{figure*}[t]
  \centering
  \includegraphics[width=\linewidth]{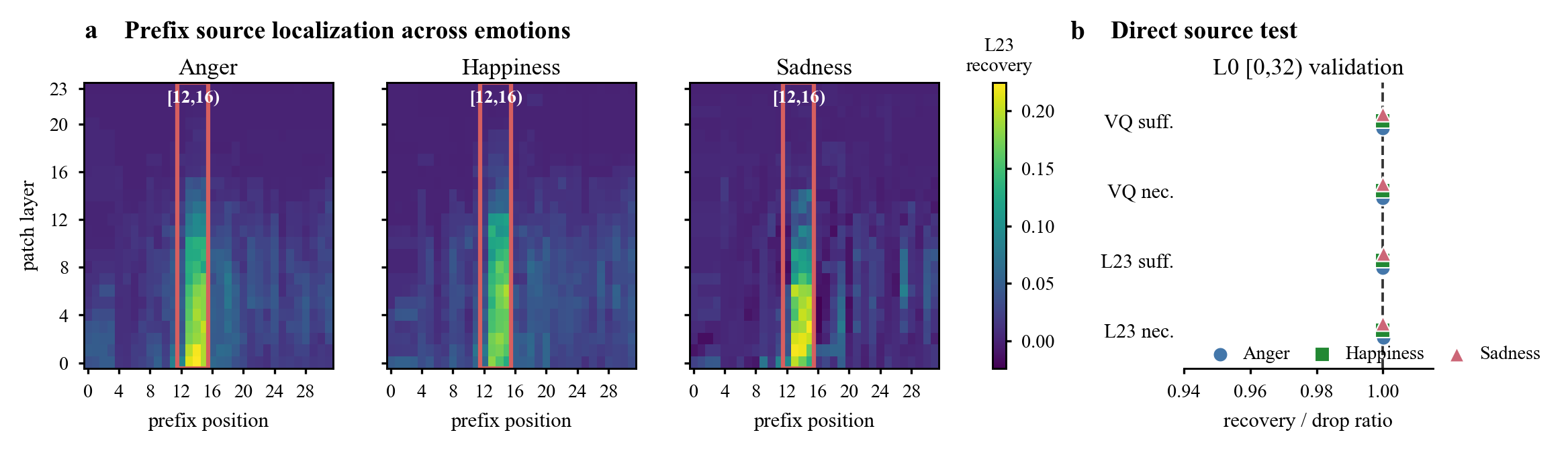}
  \caption{\textbf{Conditioning-prefix source localization and direct source tests.}
  (a) Layer-by-position scans show that emotional-prefix patching into neutral runs
  has the strongest final-layer residual recovery in the conditioning prefix, with
  a narrow peak near positions $[12,16)$.
  (b) Direct interventions on the layer-0 $[0,32)$ source window validate
  this source across anger, happiness, and sadness: emotional-prefix patching
  recovers the emotion effect, while neutral-prefix patching removes it. Ratios
  near one indicate recovery or removal of nearly the full clean neutral--emotion
  shift under both codec trajectory and late residual direction scores.}
  \label{fig:causal_prefix_source}
\end{figure*}

Both metrics are stable enough for intervention scoring
(Table~\ref{tab:emotion_metric_stability}). The codec trajectory score is less
consistent for happiness, consistent with more heterogeneous acoustic
realizations of happiness
(Appendix~\ref{sec:appendix_happy_codec_heterogeneity}). The late residual direction
score is coherent across emotions and serves as the internal readout target for
causal tracing. We use the final layer because it is closest to generated-token
prediction, although emotion directions are already separable in earlier layers
due to the conditioning prefix
(Appendix~\ref{sec:appendix_residual_direction_clustering}). These metrics do
not establish causality by themselves; they define the targets used to score the
interventions below.

\subsection{Conditioning-prefix patching for source localization}
\label{sec:causal_prefix_source}

With these metrics, source localization focuses on a natural candidate region in
IndexTTS2. The emotion reference is converted into an emotion vector and placed
in the GPT input prefix, before the text embeddings and before generated speech
tokens are produced. We therefore search the conditioning prefix rather than all
generated positions.

We test this with prefix patching. Let \(\tau^{(n)}\) and \(\tau^{(e)}\) denote
paired neutral and emotional trajectories with the same speaker and text but
different emotion references. For a conditioning-prefix source site
\(s_{\mathrm{src}}=(\ell,\mathcal W)\), where \(\mathcal W\) is a prefix window
at layer \(\ell\), we run the neutral trajectory while replacing
its prefix activations with the emotional counterparts,
{
\[
    h_{\ell,\mathcal W}^{(n)} \leftarrow h_{\ell,\mathcal W}^{(e)} ,
\]
}
and then measure whether the patched trajectory moves toward the emotional trajectory. We score this
movement with either metric defined above, \(M_k \in \{C_k,R_k\}\). Recovery is
normalized by the clean gap between the matched emotional and neutral
trajectories:
{
\[
  E_{s_{\mathrm{src}},k} =
  \frac{M_k(\tau_{s_{\mathrm{src}}}^{\mathrm{patch}})-M_k(\tau^{(n)})}
  {M_k(\tau^{(e)})-M_k(\tau^{(n)})}.
\]
}
Here \(s_{\mathrm{src}}=(\ell,\mathcal W)\) is the patched prefix site,
\(\tau_{s_{\mathrm{src}}}^{\mathrm{patch}}\) is the neutral trajectory after
replacing site \(s_{\mathrm{src}}\) with its emotional counterpart, and \(M_k\)
is either the codec trajectory score or the late residual direction score for
target emotion \(k\), using the pair-relative convention defined above. The
ratio is defined only when the magnitude of the clean gap exceeds the numerical
tolerance of \(10^{-8}\); otherwise, it is excluded. A value near one means that the patched prefix activations recover most of the clean
emotional shift under that metric. For necessity tests, we reverse the patch direction and measure how
much of the clean shift is removed when the emotional trajectory receives the neutral prefix. We use
these normalized recovery and removal scores to compare prefix sites.

\subsection{Localizing and validating the conditioning-prefix source}
\label{sec:cross_emotion_localization_controls}

In the IndexTTS2 autoregressive module, the prefix scan localizes the strongest
reference-emotion source to the conditioning prefix. Across anger,
happiness, and sadness, recovery concentrates near prefix positions \([12,16)\),
with the strongest signal in low-to-middle patch layers and weaker spread across
nearby prefix positions (Figure~\ref{fig:causal_prefix_source}a).
Direct tests on this prefix hotspot \([12,16)\) confirm that it is a strong partial source:
across emotions, \([12,16)\) recovers or removes roughly half to two thirds of
the clean shift (Appendix~\ref{sec:appendix_source_windows}).
Rather than using this narrow peak as the only source, we use the full layer-0
conditioning-prefix block \([0,32)\) as a conservative layer-0 \([0,32)\) source window for
subsequent tracing. We validate this source choice with two direct interventions
(Figure~\ref{fig:causal_prefix_source}b): patching the emotional prefix into the
neutral run should recover the emotional effect, and patching the neutral prefix
into the emotional run should remove it. Across eight matched examples per
emotion, both interventions are nearly complete under both metrics. The mean
codec recovery and removal ratios are 1.00 for all three emotions, and the late
residual direction ratios are also approximately 1.00. Thus, for this model, the
full conditioning-prefix block provides a conservative layer-0 \([0,32)\) source window, not a
claim that every token in the window is individually necessary.

These controls isolate the reference emotion from simpler confounds. Each matched pair keeps the text
and speaker fixed, so the patched contrast is not explained by lexical content or speaker identity.
The same source window is used for every validation prompt, so the result does not rely on
prompt-specific window choices. This gives a fixed source for the component analyses below.

\section{Selecting Candidate Emotion-Carrying Components}
\label{sec:candidate_components}
The localized conditioning-prefix source gives us a place to start, but not a component-level
pathway. We next use gradient attribution as a screening step to identify heads
and MLPs whose source-induced changes point toward the late emotion readout. This
yields a sparse candidate set for each emotion.

\subsection{Gradient attribution selects components}
\label{sec:attn_mlp_decomposition}

Directly intervening on every head and MLP would make candidate discovery expensive and hard to
separate from validation. We therefore use gradient attribution as a first-order screen, following
attribution patching methods for scalable component localization
\citep{nanda2023attribution,kramar2024atp}. Let
\(\tau_{s_{\mathrm{src}}}^{\mathrm{patch}}\) denote the neutral trajectory after
patching the validated conditioning-prefix source \(s_{\mathrm{src}}=(0,\mathcal W_{\mathrm{src}})\),
with \(\mathcal W_{\mathrm{src}}=[0,32)\), from the emotional trajectory. For a
component \(c\), pilot pair \(i\), and autoregressive prediction step \(t\), define the
source-induced activation change
{
\[
  \Delta a_{i,c,t} =
  a_{i,c,t}(\tau_{i,s_{\mathrm{src}}}^{\mathrm{patch}})
  - a_{i,c,t}(\tau_i^{(n)}),
\]
}
and sum its local first-order effects over prediction steps:
{\small
\[
  A_{i,c,k} = \sum_t
  \left\langle
  \Delta a_{i,c,t},
  \nabla_{a_{i,c,t}} R_k(\tau_{i,s_{\mathrm{src}}}^{\mathrm{patch}})
  \right\rangle .
\]
}
Here \(R_k\) is the late residual direction score for target emotion \(k\), and
\(a_{i,c,t}(\tau)\) denotes the generation-time output of component \(c\) for
pair \(i\) at step \(t\). The gradient is evaluated on the source-patched
trajectory. A large positive \(A_{i,c,k}\) means that the source-induced change
in component \(c\) locally increases the emotion readout. For each emotion, we
average \(A_{i,c,k}\) over eight pilot pairs and keep components whose mean score
exceeds \(\eta\bar g_k\), where \(\eta=0.005\) and
\(\bar g_k=\frac{1}{8}\sum_i\|\bar h_L(\tau_i^{(e)})-\bar h_L(\tau_i^{(n)})\|_2\)
is the mean clean pairwise final-layer residual shift over the same pilot pairs.
We also require positive attribution on at least seven pairs.
Eligible components are then ranked by mean absolute attribution, and the top
30\% form the frozen candidate set.

\begin{table}[t]
\centering
\small
\setlength{\tabcolsep}{3pt}
\renewcommand{\arraystretch}{1.08}
\begin{tabular}{@{}lccccc@{}}
\hline
\textbf{Emotion} & \textbf{Nodes} & \textbf{Heads} & \textbf{MLPs} & \shortstack{\textbf{Late}\\\textbf{MLPs}} & \textbf{Shared} \\
\hline
Anger & 24 & 15 & 9 & 6 & 13 \\
Happiness & 23 & 13 & 10 & 6 & 13 \\
Sadness & 27 & 21 & 6 & 5 & 13 \\
\hline
\end{tabular}
\caption{\textbf{Gradient-attribution candidate sets.}
Nodes are selected attention-head and MLP components in the frozen top-30\%
candidate set for each emotion. The Late MLPs column counts selected MLPs in layers 18--23.
Shared counts components that appear in all three emotion-specific candidate
sets.}
\label{tab:gradient_attribution_candidates}
\end{table}

\begin{table*}[t]
  \centering
  \small
  \begin{tabular}{lcrrrr}
  \hline
  \textbf{Emotion} & \textbf{Components} & \textbf{Pilot recover} & \textbf{Pilot remove} & \textbf{Valid. recover} & \textbf{Valid. remove} \\
  \hline
  Anger & 24 (15/9) & 0.812 / 0.815 & 0.787 / 0.786 & 0.814 / 0.815 & 0.784 / 0.787 \\
  Happiness & 23 (13/10) & 0.883 / 0.886 & 0.864 / 0.865 & 0.876 / 0.868 & 0.855 / 0.855 \\
  Sadness & 27 (21/6) & 0.751 / 0.754 & 0.754 / 0.755 & 0.736 / 0.739 & 0.747 / 0.744 \\
  \hline
  \end{tabular}
  \caption{\textbf{Pilot and validation results for frozen top-30\% component sets.}
  Each component set is selected on eight pilot pairs and evaluated without
  retuning on eight validation synthesis pairs. Components reports total selected
  components, with heads/MLPs in parentheses. Recover is the fraction of the
  full-source readout shift restored by the selected components; remove is the fraction
  lost when the same components are replaced with neutral activations in the
  full-source trajectory. Each
  recover/remove cell reports mean/median.}
  \label{tab:top30_pathway_validation}
  \end{table*}

Table~\ref{tab:gradient_attribution_candidates} shows that the attribution
ranking is sparse, late-MLP-heavy, and partly shared across emotions. The 30\% budget is also supported by
a compression sweep on anger: the top 10\% recovers 0.696 of the readout shift
and removes 0.661, the top 15\% reaches 0.763 and 0.720, and the top 30\%
reaches 0.815 and 0.798. Increasing the budget to 50\% or 90\% gives only
smaller additional gains (Appendix~\ref{sec:appendix_component_analyses}). We
therefore use the top 30\% as a compact candidate set for validation, not as a
claim that lower-ranked components have no effect.
The exact frozen component identities are reported in
Appendix~\ref{sec:appendix_component_identities}.
This gives a concrete source-to-readout candidate pathway for the intervention
tests below.

\section{Validating the Source-to-Readout Component-Level Circuit}
\label{sec:source_to_readout_pathway}
The attribution ranking gives candidate heads and MLPs, but direct interventions
are needed to test whether they carry the source effect to the emotion readout.
We therefore freeze the selected components and evaluate them on validation
synthesis cases. We test sufficiency and necessity, trace source-to-component
effects against matched controls, and compare the selected sets with matched
random component sets.
  \begin{table}[t]
    \centering
    \small
    \setlength{\tabcolsep}{4pt}
    \begin{tabular}{lcccc}
    \hline
    \textbf{Emotion} & \multicolumn{2}{c}{\textbf{Recover}} & \multicolumn{2}{c}{\textbf{Remove}} \\
    & \textbf{Selected} & \textbf{Random} & \textbf{Selected} & \textbf{Random} \\
    \hline
    Anger & 0.814 & 0.278 & 0.784 & 0.242 \\
    Happiness & 0.876 & 0.421 & 0.855 & 0.463 \\
    Sadness & 0.736 & 0.340 & 0.747 & 0.395 \\
    \hline
    \end{tabular}
    \caption{\textbf{Matched random controls for component-set validation.}
    Selected cells report validation mean ratios for the frozen component sets.
    Random cells report the mean over 20 matched random sets with the same size
    and head--MLP mix. The selected set is larger than every matched random set in
    all three emotions and both intervention directions.}
    \label{tab:matched_random_controls}
    \end{table}
    
\subsection{Frozen component interventions}
\label{sec:path_patching_setup}

We apply the sufficiency and necessity logic from source localization
(Section~\ref{sec:causal_prefix_source}) to the frozen component sets selected
above (Section~\ref{sec:candidate_components}). For each emotion, the source
window, readout metric, component budget, and selected heads and MLPs are fixed
before validation. Thus the following interventions test the
attribution-selected set directly, rather than continuing the selection
procedure.

In the sufficiency test, the neutral trajectory receives the selected head and
MLP outputs from the trajectory with the emotional source. In the necessity test,
the source-patched neutral trajectory receives the matched clean-neutral outputs
at the same components. We normalize both scores by the full-source readout gap
between the source-patched neutral and clean-neutral trajectories. A ratio near one means that the frozen
component set recovers the readout shift in the sufficiency test or removes it in
the necessity test.

\subsection{Frozen components recover and suppress the readout}
\label{sec:pathway_validation_tests}
The frozen component sets pass the main readout-level validation on synthesis
cases that were not used for attribution. Table~\ref{tab:top30_pathway_validation}
reports two ratios for each emotion: recover is the sufficiency score, and
remove is the necessity score. Across emotions, validation recovery/removal
ratios remain close to the pilot ratios, and the expected direction holds for
every pair in every condition.

Together, the two intervention directions show that the selected components both
restore the emotion readout when added to a neutral trajectory and suppress it
when replaced with clean-neutral outputs in the source-patched neutral trajectory. This supports the readout-level claim
that the frozen sparse sets carry a large fraction of the measured emotion
shift.
\subsection{Testing component specificity}
\label{sec:matched_random_component_controls}

The set-level interventions show that the selected components matter, but do not
by themselves rule out a simpler explanation: perhaps many components in similar
layers would work equally well. We therefore test specificity at the full-set
level by sampling random component sets with the same size and head/MLP mix.
This control gives a larger separation
(Table~\ref{tab:matched_random_controls}). Across emotions, selected validation
recovery ratios are 0.736--0.876, while matched random sets average 0.278--0.421.
The same pattern holds for removal: selected sets remove 0.747--0.855 of the
readout shift, while matched random sets average 0.242--0.463. In every emotion
and intervention type, the selected set exceeds all 20 random sets.

As an auxiliary source-conditioned specificity check, selected components exhibit
positive selected-minus-control deltas against cleaner matched alternatives in
all three emotions and both intervention directions
(Appendix~\ref{sec:appendix_component_specificity_controls}). Together, these
controls show that the result is not explained by patching arbitrary components
in similar layers.

\subsection{Shared and emotion-specific structure}
\label{sec:shared_specific_components}

The selected sets combine a common backbone with emotion-specific components.
The same 13 components occur in all three top sets. As
Table~\ref{tab:shared_specific_decomposition} shows, these shared components
alone recover 79--86\% of the full-set sufficiency effect, while the remaining
emotion-specific components recover 58--59\%. Both subsets have positive effects
on every held-out pair.

\begin{table}[t]
  \centering
  \small
  \caption{\textbf{Sufficiency recovery of full, shared-only, and
  emotion-specific component sets.} Parentheses give the fraction of the
  full-set recovery.}
  \label{tab:shared_specific_decomposition}
  \begin{tabular}{lccc}
    \toprule
    \textbf{Emotion} & \textbf{Full} & \textbf{Shared only} & \textbf{Specific only} \\
    \midrule
    Anger & 0.814 & 0.686 (84\%) & 0.470 (58\%) \\
    Happiness & 0.876 & 0.753 (86\%) & 0.511 (58\%) \\
    Sadness & 0.736 & 0.584 (79\%) & 0.433 (59\%) \\
    \bottomrule
  \end{tabular}
\end{table}

Shared sites nevertheless carry emotion-dependent activation content. Replacing
their target-emotion activations with activations from another emotion reduces
the target readout in 47 of 48 source--target cases. The mean changes are
\(-33.14\), \(-40.72\), and \(-51.46\) for anger, happiness, and sadness,
respectively; the overall change is \(-41.77\), with a 95\% confidence interval
of \([-48.27,-35.28]\). Thus the circuit uses shared computational sites while
retaining target-specific activation content and components.

\begin{table*}[!b]
  \centering
  \scriptsize
  \setlength{\tabcolsep}{3.5pt}
  \caption{\textbf{Decoded acoustic effects of the sparse component intervention
  and a readout-matched residual-direction baseline.} Changes are measured from
  the matched neutral baseline. The second correlate follows the salient
  target-reference contrast for each emotion.}
  \label{tab:decoded_acoustic_and_steering}
  \begin{tabular}{lccccc}
    \toprule
    \textbf{Emotion} & \textbf{Circuit \(\Delta\)F0 (st)} & \textbf{Second correlate}
    & \textbf{Circuit change} & \textbf{Residual \(\Delta\)F0} & \textbf{Ratio} \\
    \midrule
    Anger & \(+2.77\) \([2.44,3.12]\) & RMS energy
      & \(+3.35\) dB \([2.99,3.73]\) & \(+0.74\) st & 0.27 \\
    Happiness & \(+3.57\) \([3.18,3.98]\) & Spectral centroid
      & \(+133.06\) Hz \([113.37,152.10]\) & \(+0.83\) st & 0.23 \\
    Sadness & \(-1.01\) \([-1.38,-0.55]\) & RMS energy
      & \(-3.04\) dB \([-3.42,-2.67]\) & \(-0.17\) st & 0.17 \\
    \bottomrule
  \end{tabular}
\end{table*}

\section{Decoded-Audio Effects and Baselines}
\label{sec:audio_level_top30_validation}

The component tests so far are internal; we next ask whether the same fixed
interventions affect decoded waveforms. For each target emotion, we compare 24
matched text-speaker cases under a neutral baseline and a component-patched
target intervention. We measure mean F0, active-frame RMS energy, and spectral
centroid, calibrating each sign by the corresponding target-minus-neutral
reference contrast. We report paired bootstrap 95\% confidence intervals,
paired sign-permutation tests, and directional consistency. Target-prototype
Emo-SIM from emotion2vec \citep{ma-etal-2024-emotion2vec} is reported as a
secondary measure in Appendix~\ref{sec:appendix_decoded_audio_stats}.

\subsection{Paired acoustic effects}
Table~\ref{tab:decoded_acoustic_and_steering} shows consistent waveform-level
changes for all three emotions. Anger increases pitch and energy, happiness
increases pitch and spectral brightness, and sadness decreases pitch and energy,
matching the corresponding reference contrasts. All reported acoustic changes
are significant under paired permutation tests: \(p\leq 10^{-4}\) with 24/24
directional consistency, except sadness F0 with \(p=3\times10^{-4}\) and 23/24
consistency.

\subsection{Readout-matched residual-direction steering}
We compare the component intervention with direct steering along the layer-23
residual emotion direction at every generation step. Steering strength is chosen
to match the mean component-induced readout shift. The residual baseline reaches
equal or larger internal shifts---56.9 versus 51.0 for anger, 62.1 versus 57.5
for happiness, and 35.1 versus 32.4 for sadness---but produces only 17--27\% of
the component intervention's F0 change (Table~\ref{tab:decoded_acoustic_and_steering}).
The selected components therefore transfer reference-derived emotion information
to decoded acoustics more effectively than a global readout direction.

The effect also persists under top-\(p\) sampling: over eight pairs per emotion,
F0 changes by \(+2.15\), \(+1.69\), and \(-0.95\) semitones for anger,
happiness, and sadness. A second-backbone check on LLaSA-1B further finds that a
layer-8 anger-prefix patch recovers 0.855 of the speech-token direction effect
over 30 prompts. Appendix~\ref{sec:appendix_robustness_checks} provides the full
protocols and confidence intervals. An exploratory blinded A/B check also
prefers the intervention for target-emotion strength in 24/24 trials, while 21
of 24 naturalness comparisons are ties
(Appendix~\ref{sec:appendix_listening_check}).

\section{Conclusion}
\label{sec:conclusion}
This work traces reference-conditioned emotion from an input-prefix source
through sparse attention-head and MLP sets to late readouts and decoded speech.
The selected component-level circuits recover and suppress most of the measured
readout effect on held-out cases, combine a shared computational backbone with
emotion-specific content, and outperform matched random sets. Their decoded
acoustic effects are substantially stronger than those of readout-matched
residual-direction steering, linking the localized components to concrete
changes in pitch, energy, and spectral brightness. These results establish a
compact causal account of reference-emotion routing in LLM-based TTS under
matched-reference conditions.

\section*{Limitations}
\label{sec:limitations}
The main circuit analysis uses IndexTTS2, one speaker/timbre setting, English
prompts, three emotions, and one fixed reference file per emotion. The identified
components may therefore capture properties of these references in addition to
emotion structure. The LLaSA-1B and top-\(p\) checks extend the causal pattern to
a second backbone and stochastic decoding, but broader tests across speakers,
languages, references, and architectures remain open. Our circuit is a
component-level source-to-readout account rather than a complete enumeration of
every internal edge. Finally, the listening check uses one listener; a
multi-listener study would provide stronger perceptual evaluation.

\section*{Ethical Considerations and Potential Risks}
\label{sec:ethics_risks}
Emotion-controllable TTS could be misused to misrepresent a speaker's affect or
to create deceptive voice content, especially when combined with unauthorized
voice cloning. Our work is a controlled interpretability study and does not
release a deployable editing system. Any deployment of emotion-controllable TTS
should require speaker consent, clear disclosure or provenance mechanisms, and
safeguards against impersonation and deceptive use.

\section*{Acknowledgments}
This work is supported in part by the National Science Foundation (NSF) grant
IIS-2451436 and the Commonwealth Cyber Initiative grant HC-2Q26-032.

\bibliography{custom}

@article{wang2023neural,
  title = {Neural Codec Language Models are Zero-Shot Text to Speech Synthesizers},
  author = {Wang, Chengyi and Chen, Sanyuan and Wu, Yu and Zhang, Ziqiang and Zhou, Long and Liu, Shujie and Chen, Zhuo and Liu, Yanqing and Wang, Huaming and Li, Jinyu and He, Lei and Zhao, Sheng and Wei, Furu},
  journal = {arXiv preprint arXiv:2301.02111},
  year = {2023},
  url = {https://arxiv.org/abs/2301.02111}
}

@article{borsos2023audiolm,
  title = {{AudioLM}: A Language Modeling Approach to Audio Generation},
  author = {Borsos, Zal{\'a}n and Marinier, Rapha{\"e}l and Vincent, Damien and Kharitonov, Eugene and Pietquin, Olivier and Sharifi, Matt and Roblek, Dominik and Teboul, Olivier and Grangier, David and Tagliasacchi, Marco and Zeghidour, Neil},
  journal = {IEEE/ACM Transactions on Audio, Speech, and Language Processing},
  volume = {31},
  pages = {2523--2533},
  year = {2023},
  doi = {10.1109/TASLP.2023.3288409}
}

@inproceedings{guo2022prompttts,
  title = {{PromptTTS}: Controllable Text-to-Speech with Text Descriptions},
  author = {Guo, Zhifang and Leng, Yichong and Wu, Yihan and Zhao, Sheng and Tan, Xu},
  booktitle = {2023 IEEE International Conference on Acoustics, Speech and Signal Processing (ICASSP)},
  pages = {1--5},
  year = {2023},
  publisher = {IEEE},
  doi = {10.1109/ICASSP49357.2023.10096285}
}

@article{yang2023instructtts,
  title = {{InstructTTS}: Modelling Expressive {TTS} in Discrete Latent Space With Natural Language Style Prompt},
  author = {Yang, Dongchao and Liu, Songxiang and Huang, Rongjie and Weng, Chao and Meng, Helen},
  journal = {IEEE/ACM Transactions on Audio, Speech, and Language Processing},
  volume = {32},
  pages = {2913--2925},
  year = {2024},
  publisher = {IEEE},
  doi = {10.1109/TASLP.2024.3402088}
}

@article{du2024cosyvoice,
  title = {{CosyVoice}: A Scalable Multilingual Zero-Shot Text-to-Speech Synthesizer Based on Supervised Semantic Tokens},
  author = {Du, Zhihao and Chen, Qian and Zhang, Shiliang and Hu, Kai and Lu, Heng and Yang, Yexin and Hu, Hangrui and Zheng, Siqi and Gu, Yue and Ma, Ziyang and Gao, Zhifu and Yan, Zhijie},
  journal = {arXiv preprint arXiv:2407.05407},
  year = {2024},
  url = {https://arxiv.org/abs/2407.05407}
}

@article{wang2025sparktts,
  title = {{Spark-TTS}: An Efficient {LLM}-Based Text-to-Speech Model with Single-Stream Decoupled Speech Tokens},
  author = {Wang, Xinsheng and Jiang, Mingqi and Ma, Ziyang and Zhang, Ziyu and Liu, Songxiang and Li, Linqin and Liang, Zheng and Zheng, Qixi and Wang, Rui and Feng, Xiaoqin and Bian, Weizhen and Ye, Zhen and Cheng, Sitong and Yuan, Ruibin and Zhao, Zhixian and Zhu, Xinfa and Pan, Jiahao and Xue, Liumeng and Zhu, Pengcheng and Chen, Yunlin and Li, Zhifei and Chen, Xie and Xie, Lei and Guo, Yike and Xue, Wei},
  journal = {arXiv preprint arXiv:2503.01710},
  year = {2025},
  url = {https://arxiv.org/abs/2503.01710}
}

@book{picard1997affective,
  title = {Affective Computing},
  author = {Picard, Rosalind W.},
  publisher = {MIT Press},
  address = {Cambridge, MA},
  year = {1997},
  isbn = {978-0-262-16170-1},
  url = {https://mitpress.mit.edu/9780262161701/affective-computing/}
}

@article{bickmore2005establishing,
  title = {Establishing and Maintaining Long-Term Human-Computer Relationships},
  author = {Bickmore, Timothy W. and Picard, Rosalind W.},
  journal = {ACM Transactions on Computer-Human Interaction},
  volume = {12},
  number = {2},
  pages = {293--327},
  year = {2005},
  doi = {10.1145/1067860.1067867}
}

@article{scherer2003vocal,
  title = {Vocal Communication of Emotion: A Review of Research Paradigms},
  author = {Scherer, Klaus R.},
  journal = {Speech Communication},
  volume = {40},
  number = {1--2},
  pages = {227--256},
  year = {2003},
  doi = {10.1016/S0167-6393(02)00084-5}
}

@article{juslin2003communication,
  title = {Communication of Emotions in Vocal Expression and Music Performance: Different Channels, Same Code?},
  author = {Juslin, Patrik N. and Laukka, Petri},
  journal = {Psychological Bulletin},
  volume = {129},
  number = {5},
  pages = {770--814},
  year = {2003},
  doi = {10.1037/0033-2909.129.5.770}
}

@article{olah2020zoom,
  title = {Zoom In: An Introduction to Circuits},
  author = {Olah, Chris and Cammarata, Nick and Schubert, Ludwig and Goh, Gabriel and Petrov, Michael and Carter, Shan},
  journal = {Distill},
  year = {2020},
  doi = {10.23915/distill.00024.001},
  url = {https://distill.pub/2020/circuits/zoom-in/}
}

@article{elhage2021mathematical,
  title = {A Mathematical Framework for Transformer Circuits},
  author = {Elhage, Nelson and Nanda, Neel and Olsson, Catherine and Henighan, Tom and Joseph, Nicholas and Mann, Ben and Askell, Amanda and Bai, Yuntao and Chen, Anna and Conerly, Tom and DasSarma, Nova and Drain, Dawn and Ganguli, Deep and Hatfield-Dodds, Zac and Hernandez, Danny and Jones, Andy and Kernion, Jackson and Lovitt, Liane and Ndousse, Kamal and Amodei, Dario and Brown, Tom and Clark, Jack and Kaplan, Jared and McCandlish, Sam and Olah, Chris},
  journal = {Transformer Circuits Thread},
  year = {2021},
  url = {https://transformer-circuits.pub/2021/framework/index.html}
}

@inproceedings{meng2022locating,
  title = {Locating and Editing Factual Associations in {GPT}},
  author = {Meng, Kevin and Bau, David and Andonian, Alex and Belinkov, Yonatan},
  booktitle = {Advances in Neural Information Processing Systems},
  volume = {35},
  pages = {17359--17372},
  year = {2022},
  doi = {10.52202/068431-1262},
  url = {https://proceedings.neurips.cc/paper_files/paper/2022/hash/6f1d43d5a82a37e89b0665b33bf3a182-Abstract-Conference.html}
}

@article{conmy2023automated,
  title = {Towards Automated Circuit Discovery for Mechanistic Interpretability},
  author = {Conmy, Arthur and Mavor-Parker, Augustine and Lynch, Aengus and Heimersheim, Stefan and Garriga-Alonso, Adri{\`a}},
  journal = {Advances in Neural Information Processing Systems},
  volume = {36},
  pages = {16318--16352},
  year = {2023},
  doi = {10.52202/075280-0719},
  url = {https://proceedings.neurips.cc/paper_files/paper/2023/hash/34e1dbe95d34d7ebaf99b9bcaeb5b2be-Abstract-Conference.html}
}

@misc{olsson2022induction,
  title = {In-context Learning and Induction Heads},
  author = {Olsson, Catherine and Elhage, Nelson and Nanda, Neel and Joseph, Nicholas and DasSarma, Nova and Henighan, Tom and Mann, Ben and Askell, Amanda and Bai, Yuntao and Chen, Anna and Conerly, Tom and Drain, Dawn and Ganguli, Deep and Hatfield-Dodds, Zac and Hernandez, Danny and Johnston, Scott and Jones, Andy and Kernion, Jackson and Lovitt, Liane and Ndousse, Kamal and Amodei, Dario and Brown, Tom and Clark, Jack and Kaplan, Jared and McCandlish, Sam and Olah, Chris},
  year = {2022},
  howpublished = {Transformer Circuits Thread},
  url = {https://transformer-circuits.pub/2022/in-context-learning-and-induction-heads/index.html}
}

@inproceedings{wang2023interpretability,
  title = {Interpretability in the Wild: a Circuit for Indirect Object Identification in {GPT}-2 Small},
  author = {Wang, Kevin Ro and Variengien, Alexandre and Conmy, Arthur and Shlegeris, Buck and Steinhardt, Jacob},
  booktitle = {International Conference on Learning Representations},
  year = {2023},
  url = {https://openreview.net/forum?id=NpsVSN6o4ul}
}

@inproceedings{vig2020causal,
  title = {Investigating Gender Bias in Language Models Using Causal Mediation Analysis},
  author = {Vig, Jesse and Gehrmann, Sebastian and Belinkov, Yonatan and Qian, Sharon and Nevo, Daniel and Singer, Yaron and Shieber, Stuart},
  booktitle = {Advances in Neural Information Processing Systems},
  volume = {33},
  pages = {12388--12401},
  year = {2020},
  url = {https://proceedings.neurips.cc/paper/2020/hash/92650b2e92217715fe312e6fa7b90d82-Abstract.html}
}

@article{heimersheim2024activation,
  title = {How to Use and Interpret Activation Patching},
  author = {Heimersheim, Stefan and Nanda, Neel},
  journal = {arXiv preprint arXiv:2404.15255},
  year = {2024},
  url = {https://arxiv.org/abs/2404.15255}
}

@inproceedings{glazer2026beyond,
  title = {Beyond Transcription: Mechanistic Interpretability in {ASR}},
  author = {Glazer, Neta and Segal-Feldman, Yael and Segev, Hilit and Shamsian, Aviv and Buchnick, Asaf and Hetz, Gill and Fetaya, Ethan and Keshet, Joseph and Navon, Aviv},
  booktitle = {Proceedings of the AAAI Conference on Artificial Intelligence},
  volume = {40},
  pages = {37407--37416},
  year = {2026},
  doi = {10.1609/aaai.v40i44.41073},
  url = {https://ojs.aaai.org/index.php/AAAI/article/view/41073}
}

@article{goldowskydill2023path,
  title = {Localizing Model Behavior with Path Patching},
  author = {Goldowsky-Dill, Nicholas and MacLeod, Chris and Sato, Lucas and Arora, Aryaman},
  journal = {arXiv preprint arXiv:2304.05969},
  year = {2023},
  url = {https://arxiv.org/abs/2304.05969}
}

@misc{nanda2023attribution,
  title = {Attribution Patching: Activation Patching at Industrial Scale},
  author = {Nanda, Neel},
  year = {2023},
  howpublished = {Online article},
  url = {https://www.neelnanda.io/mechanistic-interpretability/attribution-patching},
  note = {Accessed: 2026-08-25}
}

@article{kramar2024atp,
  title = {{AtP*}: An Efficient and Scalable Method for Localizing {LLM} Behaviour to Components},
  author = {Kram{\'a}r, J{\'a}nos and Lieberum, Tom and Shah, Rohin and Nanda, Neel},
  journal = {arXiv preprint arXiv:2403.00745},
  year = {2024},
  url = {https://arxiv.org/abs/2403.00745}
}

@article{zhang2025proemo,
  title = {{PROEMO}: Prompt-Driven Text-to-Speech Synthesis Based on Emotion and Intensity Control},
  author = {Zhang, Shaozuo and Mehrish, Ambuj and Li, Yingting and Poria, Soujanya},
  journal = {arXiv preprint arXiv:2501.06276},
  year = {2025},
  url = {https://arxiv.org/abs/2501.06276}
}

@article{deng2025indextts,
  title = {{IndexTTS}: An Industrial-Level Controllable and Efficient Zero-Shot Text-to-Speech System},
  author = {Deng, Wei and Zhou, Siyi and Shu, Jingchen and Wang, Jinchao and Wang, Lu},
  journal = {arXiv preprint arXiv:2502.05512},
  year = {2025},
  url = {https://arxiv.org/abs/2502.05512}
}

@inproceedings{zhou2025indextts2,
  title = {{IndexTTS2}: A Breakthrough in Emotionally Expressive and Duration-Controlled Auto-Regressive Zero-Shot Text-to-Speech},
  author = {Zhou, Siyi and Zhou, Yiquan and He, Yi and Zhou, Xun and Wang, Jinchao and Deng, Wei and Shu, Jingchen},
  booktitle = {Proceedings of the AAAI Conference on Artificial Intelligence},
  volume = {40},
  pages = {35139--35148},
  year = {2026},
  doi = {10.1609/aaai.v40i41.40820},
  url = {https://ojs.aaai.org/index.php/AAAI/article/view/40820}
}

@inproceedings{yang2025emovoice,
  title = {{EmoVoice}: {LLM}-Based Emotional Text-to-Speech Model with Freestyle Text Prompting},
  author = {Yang, Guanrou and Yang, Chen and Chen, Qian and Ma, Ziyang and Chen, Wenxi and Wang, Wen and Wang, Tianrui and Yang, Yifan and Niu, Zhikang and Liu, Wenrui and Yu, Fan and Du, Zhihao and Gao, Zhifu and Zhang, Shiliang and Chen, Xie},
  booktitle = {Proceedings of the 33rd ACM International Conference on Multimedia},
  pages = {10748--10757},
  year = {2025},
  publisher = {Association for Computing Machinery},
  doi = {10.1145/3746027.3754829}
}

@inproceedings{li2026emorltts,
  title = {{EMORL-TTS}: Reinforcement Learning for Fine-Grained Emotion Control in {LLM}-Based {TTS}},
  author = {Li, Haoxun and Liu, Yu and Sun, Yuqing and Shi, Hanlei and Qu, Leyuan and Li, Taihao},
  booktitle = {2026 IEEE International Conference on Acoustics, Speech and Signal Processing (ICASSP)},
  pages = {17272--17276},
  year = {2026},
  publisher = {IEEE},
  doi = {10.1109/ICASSP55912.2026.11460999}
}

@article{paek2025learning,
  title = {Learning Interpretable Features in Audio Latent Spaces via Sparse Autoencoders},
  author = {Paek, Nathan and Zang, Yongyi and Yang, Qihui and Leistikow, Randal},
  journal = {arXiv preprint arXiv:2510.23802},
  year = {2025},
  url = {https://arxiv.org/abs/2510.23802}
}

@article{xie2025emosteer,
  title = {{EmoSteer-TTS}: Fine-Grained and Training-Free Emotion-Controllable Text-to-Speech via Activation Steering},
  author = {Xie, Tianxin and Yang, Shan and Li, Chenxing and Yu, Dong and Liu, Li},
  journal = {arXiv preprint arXiv:2508.03543},
  year = {2025},
  url = {https://arxiv.org/abs/2508.03543}
}

@inproceedings{ma-etal-2024-emotion2vec,
  title = {emotion2vec: Self-Supervised Pre-Training for Speech Emotion Representation},
  author = {Ma, Ziyang and Zheng, Zhisheng and Ye, Jiaxin and Li, Jinchao and Gao, Zhifu and Zhang, ShiLiang and Chen, Xie},
  booktitle = {Findings of the Association for Computational Linguistics: ACL 2024},
  pages = {15747--15760},
  year = {2024},
  address = {Bangkok, Thailand},
  publisher = {Association for Computational Linguistics},
  doi = {10.18653/v1/2024.findings-acl.931},
  url = {https://aclanthology.org/2024.findings-acl.931/}
}

\appendix

\section{Experimental Setup Details}
\label{sec:appendix_experimental_setup}

\paragraph{Model and tracing configuration.}
All experiments use the released IndexTTS2-style autoregressive GPT/T2S module
with the checkpoint configuration loaded from \texttt{index-tts/checkpoints}.
The model is run in evaluation mode, and the accelerated inference engine is
disabled during tracing so that PyTorch hooks can capture and patch internal
states. Unless otherwise noted, generation traces are capped at 200 generated
speech token steps.

\paragraph{Text and reference manifests.}
The paired manifests used in our experiments contain 50 English text prompts
and four reference conditions per prompt: neutral, anger, happiness, and
sadness, for 200 rows in each manifest. They use a single speaker identifier,
\texttt{0001}. Within each neutral--emotion pair, the text and speaker/timbre
reference are fixed, while the style reference is an utterance expressing the
target emotion. The exact style-reference utterance used for each run is
specified in its corresponding manifest.

\paragraph{Pair construction.}
For each text id \(i\) and target emotion \(k \in \{\mathrm{anger},
\mathrm{happy}, \mathrm{sad}\}\), we construct a pair identifier such as
\texttt{text0\_anger} consisting of a neutral trajectory and an emotional
trajectory for the same prompt. This yields 150 possible neutral--emotion pairs:
50 text prompts for each of three non-neutral target emotions. Pair construction requires both the
neutral row and the target-emotion row to exist in the manifest and requires the
corresponding cached trace to be present. Pairs that do not satisfy these
conditions are excluded before analysis.

\paragraph{Generation and random seeds.}
The main experiments use deterministic greedy codec-token decoding. We set the
Python, NumPy, PyTorch, and CUDA random seeds to 42 in the patching scripts. A
separate robustness run uses top-\(p\) sampling with temperature 0.9 and
\(p=0.95\), holding the sampling seed fixed across intervention roles within
each matched pair.

\paragraph{Activation caches.}
For each pair, the cached trace stores both neutral and emotional trajectories.
The main cache captures residual streams, MLP outputs, attention-head outputs,
and full prefill activations for all Transformer layers. Generation-time caches
store the activation used to predict each next codec token, along with metadata
including prefix length, generated token id, and generation length. Prefix
source interventions capture donor prefill residual states from the emotional
prompt and patch them into the corresponding neutral prompt at the same layer
and prefix positions.

\paragraph{Pair regimes and validation split.}
We use two pair regimes. First, coarse source-localization scans are run over
the full 50 pairs per target emotion when available; the merged full-scan outputs
record \texttt{num\_pairs\_used=50}. Second, higher-cost interventions use fixed
8-pair pilot sets for each target emotion. The pilot ids are \texttt{text0},
\texttt{text1}, and \texttt{text10}--\texttt{text15}, paired with the target
emotion suffix. Components are selected on these eight pilot prompts and
evaluated without retuning on eight validation synthesis cases. The matched
random component controls use the same validation synthesis cases and the same
sufficiency and necessity metrics. The source-conditioned component specificity
control also uses the validation synthesis cases and keeps the frozen component
membership fixed. Decoded-audio analyses use 24 matched synthesis pairs per
emotion, and the top-\(p\) check uses eight pairs per emotion.

\section{Metric Validation and Readout Diagnostics}
\label{sec:appendix_additional_evidence}

\subsection{Happiness codec-direction heterogeneity}
\label{sec:appendix_happy_codec_heterogeneity}

Happiness has lower pair-level codec agreement than anger and sadness in
Table~\ref{tab:emotion_metric_stability}. To check whether this is driven by a few anomalous pairs, we
remove the most inconsistent happy pairs and recompute the codec-direction statistics.
Table~\ref{tab:happy_vq_outlier} shows that pairwise agreement
increases only modestly, while split-half agreement stays near 0.99, suggesting a reproducible but
more heterogeneous happiness direction.

\begin{table}[t]
\centering
\small
\begin{tabular}{ccc}
\hline
\textbf{Removed} & \textbf{Pair cos.} & \textbf{Split cos.} \\
\hline
0  & 0.764 & 0.988 \\
1  & 0.769 & 0.986 \\
3  & 0.780 & 0.988 \\
5  & 0.789 & 0.987 \\
10 & 0.810 & 0.987 \\
\hline
\end{tabular}
\caption{\textbf{Outlier removal for the Happiness codec direction.}
Removing the least aligned happiness pairs modestly increases pairwise agreement but leaves split-half
agreement high.}
\label{tab:happy_vq_outlier}
\end{table}

\subsection{Residual direction clustering}
\label{sec:appendix_residual_direction_clustering}

To examine when reference-conditioned emotion becomes separable, we visualize
paired residual differences at the first autoregressive prediction step across
layers.

\begin{figure*}[!t]
      \centering
      \includegraphics[width=\linewidth]{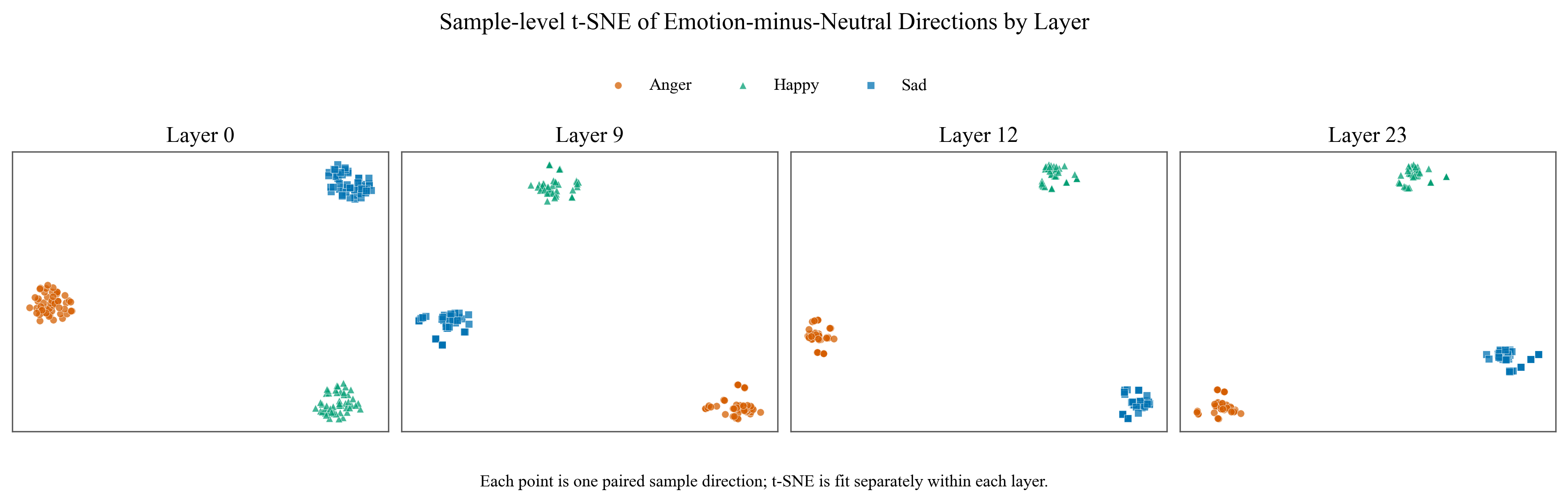}
      \caption{\textbf{Emotion clustering at the first autoregressive prediction step.}
      Each point is a paired sample-level emotion direction, computed as the residual difference
      between an emotional trajectory and its neutral counterpart at the same layer and prediction
      step. t-SNE is fit independently within each layer. This figure measures layerwise separability
      at a single prediction step; the whole-generation residual readout used for component selection
      and validation is computed separately.}
      \label{fig:emotion_direction_tsne}
\end{figure*}
\FloatBarrier
Figure~\ref{fig:emotion_direction_tsne} shows the resulting first-step
layerwise separation; the whole-generation residual readout is evaluated
separately.

\section{Decoded-Audio Statistical Details}
\label{sec:appendix_decoded_audio_stats}

We retain calibrated target-prototype Emo-SIM as an auxiliary decoded-audio
measure. Table~\ref{tab:paired_emosim_stats} reports paired bootstrap intervals,
win counts, and sign-permutation tests over the same 24 pairs per emotion used in
the acoustic analysis. Happiness and sadness show positive calibrated changes;
the anger interval includes zero. The primary decoded-audio evidence in
Section~\ref{sec:audio_level_top30_validation} is therefore the more direct and
consistent acoustic analysis.

\begin{table}[t]
  \centering
  \scriptsize
  \setlength{\tabcolsep}{3.5pt}
  \caption{\textbf{Paired statistics for calibrated Emo-SIM change.}}
  \label{tab:paired_emosim_stats}
  \begin{tabular}{lccc}
    \toprule
    \textbf{Emotion} & \textbf{Mean [95\% CI]} & \textbf{Wins} & \textbf{\(p\)} \\
    \midrule
    Anger & \(+0.323\) \([-0.209,0.833]\) & 17/24 & 0.243 \\
    Happiness & \(+0.730\) \([0.438,1.047]\) & 20/24 & \(2\times10^{-5}\) \\
    Sadness & \(+0.396\) \([0.105,0.731]\) & 19/24 & 0.0156 \\
    \bottomrule
  \end{tabular}
\end{table}

\section{Decoding and Second-Backbone Robustness}
\label{sec:appendix_robustness_checks}

\paragraph{Top-\(p\) decoding.}
We rerun the frozen component interventions with temperature 0.9 and
top-\(p=0.95\) on eight pairs per emotion. The sampling seed is shared across
roles within each pair. The resulting F0 changes preserve the greedy-decoding
directions (Table~\ref{tab:topp_robustness}), and the target readout moves in the
intended direction in all 24 cases.

\begin{table}[t]
  \centering
  \scriptsize
  \setlength{\tabcolsep}{3.5pt}
  \caption{\textbf{F0 effects under top-\(p\) decoding.} Intervals are paired
  bootstrap 95\% confidence intervals over eight pairs per emotion.}
  \label{tab:topp_robustness}
  \begin{tabular}{lc}
    \toprule
    \textbf{Emotion} & \textbf{\(\Delta\)F0 in semitones [95\% CI]} \\
    \midrule
    Anger & \(+2.15\) \([1.29,3.00]\) \\
    Happiness & \(+1.69\) \([1.20,2.22]\) \\
    Sadness & \(-0.95\) \([-1.46,-0.45]\) \\
    \bottomrule
  \end{tabular}
\end{table}

\paragraph{LLaSA-1B source localization.}
Component identities are architecture-specific, so the second-backbone test
repeats the source-sufficiency question rather than transferring IndexTTS2
components. On 30 matched anger--neutral prompts, a right-aligned layer-8 LLaSA
prefix patch recovers 0.855 of the generated speech-token anger-direction effect
(95\% CI \([0.771,0.931]\); 30/30 directional consistency). A neutral
random-window control is near zero at \(-0.044\) (95\% CI
\([-0.148,0.043]\)).

\Needspace{0.50\textheight}
\section{Exploratory Listening Check}
\label{sec:appendix_listening_check}

One of the authors completed a blinded A/B comparison over 24 matched trials,
with eight trials per target emotion. This author-conducted evaluation is
intended only as an exploratory sanity check; no external participants were
recruited, and the results should not be interpreted as population-level
perceptual evidence. Each trial compared the neutral baseline with the sparse
component intervention using the same text and reference setup. The intervention
was preferred for target-emotion strength in all 24 trials; naturalness was
judged as a tie in 21 trials
(Table~\ref{tab:single_listener_check}).

\begin{table}[H]
  \centering
  \scriptsize
  \setlength{\tabcolsep}{3pt}
  \caption{\textbf{Single-listener blinded A/B results.} Naturalness cells report
  intervention/baseline/tie counts. The reported \(p\)-values are item-level
  one-sided exact sign tests.}
  \label{tab:single_listener_check}
  \begin{tabular}{lccc}
    \toprule
    \textbf{Emotion} & \textbf{Emotion pref.} & \textbf{\(p\)} & \textbf{Naturalness} \\
    \midrule
    Anger & 8/8 & 0.0039 & 0/1/7 \\
    Happiness & 8/8 & 0.0039 & 1/1/6 \\
    Sadness & 8/8 & 0.0039 & 0/0/8 \\
    Overall & 24/24 & \(5.96\times10^{-8}\) & 1/2/21 \\
    \bottomrule
  \end{tabular}
\end{table}

\Needspace{0.52\textheight}
\section{Conditioning-Prefix Source Localization Details}
\label{sec:appendix_source_windows}

The conditioning-prefix source scan evaluates eight non-overlapping four-token prefix windows from \([0,4)\) to
\([28,32)\).
The narrow hotspot \([12,16)\) is selected from these scans. The broader
layer-0 \([0,32)\) source window is then used for downstream validation. Source-window
validation records the neutral and emotional prefix lengths and skips a pair if
the selected source window exceeds either prefix length. In the current
validation set, the \([0,32)\) envelope fits all reported pairs.
Table~\ref{tab:hotspot_source_window_tests} reports the corresponding narrow-window
sufficiency and necessity results.

\begin{table}[H]
\centering
\scriptsize
\setlength{\tabcolsep}{3pt}
\begin{tabular}{lcccc}
\hline
\textbf{Emotion} & \textbf{Codec suff.} & \textbf{Resid. suff.} & \textbf{Codec nec.} & \textbf{Resid. nec.} \\
\hline
Anger & 0.62 & 0.60 & 0.47 & 0.54 \\
Happiness & 0.55 & 0.55 & 0.73 & 0.65 \\
Sadness & 0.59 & 0.53 & 0.59 & 0.64 \\
\hline
\end{tabular}
\caption{\textbf{Direct tests on the narrow prefix hotspot \([12,16)\).}
Ratios are averaged over 50 matched pairs per emotion. The hotspot recovers or
removes a substantial fraction of the clean neutral--emotion shift, but does not
fully account for the effect; the main tracing experiments therefore use the
broader layer-0 \([0,32)\) conditioning-prefix block as the source window.}
\label{tab:hotspot_source_window_tests}
\end{table}

\section{Component Selection and Set-Level Validation Details}
\label{sec:appendix_component_analyses}

\paragraph{Component-set selection and validation boundary.}
Component candidates are selected from a source-conditioned attribution graph
using the fixed layer-0 \([0,32)\) source window and a fixed layer-23
residual emotion readout. Canonical top-30\% sets are frozen before the direct
sufficiency and necessity tests. The resulting sets contain 24 nodes for anger,
27 nodes for sadness, and 23 nodes for the happiness configuration used in the
cross-emotion component-set summary. These validation tests therefore evaluate
fixed component sets rather than retuning membership during the intervention
trajectory.

\subsection{Canonical Component Identities}
\label{sec:appendix_component_identities}

The evaluated component universe contains 480 attention heads and 24 MLPs
across 24 Transformer layers, for 504 components in total. Table~\ref{tab:canonical_component_identities}
lists the three frozen sets in mean-absolute-attribution rank order. We write
\(LxHy\) for attention head \(y\) in layer \(x\), and \(Lx\mathrm{MLP}\) for
the MLP in layer \(x\). The sets aggregate attribution across the eight pilot
pairs and are not reselected on validation pairs; attribution magnitudes and
within-set ranks may vary across individual pairs. Component identities are
specific to the studied checkpoint and reference configuration.

\begin{table}[t]
  \centering
  \scriptsize
  \setlength{\tabcolsep}{7pt}
  \renewcommand{\arraystretch}{0.92}
  \caption{\textbf{Frozen canonical component identities.} Each column follows
  its emotion-specific attribution ranking. Bold entries are the 13 components
  shared by all three sets.}
  \label{tab:canonical_component_identities}
  \begin{tabular}{lll}
    \toprule
    \textbf{Anger} & \textbf{Happiness} & \textbf{Sadness} \\
    \midrule
    \textbf{L21MLP} & \textbf{L22MLP} & L0H2 \\
    \textbf{L22MLP} & \textbf{L21MLP} & \textbf{L21MLP} \\
    \textbf{L20MLP} & L0H2 & \textbf{L22MLP} \\
    L18MLP & \textbf{L20MLP} & \textbf{L20MLP} \\
    \textbf{L16H8} & \textbf{L19MLP} & L0MLP \\
    L17MLP & \textbf{L6H2} & L0H16 \\
    \textbf{L19MLP} & L0MLP & \textbf{L6H2} \\
    \textbf{L6H2} & \textbf{L16H8} & \textbf{L16H8} \\
    \textbf{L9H14} & \textbf{L23MLP} & L9H19 \\
    \textbf{L15H19} & L17MLP & \textbf{L19MLP} \\
    \textbf{L12H19} & L18MLP & L12H0 \\
    \textbf{L23MLP} & \textbf{L9H14} & \textbf{L15H19} \\
    L16MLP & L0H16 & \textbf{L23MLP} \\
    L12H0 & \textbf{L14H13} & L11H9 \\
    L8H12 & \textbf{L10H19} & L12H9 \\
    \textbf{L14H13} & \textbf{L12H19} & L9H12 \\
    L15MLP & L12H9 & L13H13 \\
    L14H19 & \textbf{L11H10} & \textbf{L12H19} \\
    \textbf{L11H10} & \textbf{L15H19} & \textbf{L10H19} \\
    \textbf{L10H19} & L10H4 & L10H12 \\
    L14H18 & L5MLP & \textbf{L11H10} \\
    L10H12 & L6MLP & L11H14 \\
    L13H13 & L14H18 & \textbf{L9H14} \\
    L10H4 & & L10H16 \\
    & & L14H4 \\
    & & L8H14 \\
    & & \textbf{L14H13} \\
    \bottomrule
  \end{tabular}
\end{table}

\begin{figure*}[t]
  \centering
  \includegraphics[width=0.8\linewidth]{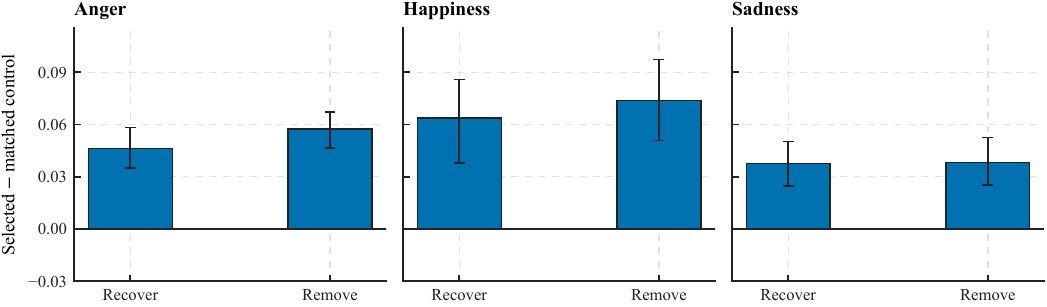}
  \caption{\textbf{Source-conditioned component specificity controls.}
  Each selected component is compared with five cleaner matched alternatives:
  same-layer non-selected heads for head components, and nearby same-type
  non-selected MLPs for MLP components. Bars report pair-level
  selected-minus-control deltas after averaging over selected components within
  each validation pair; error bars are bootstrap confidence intervals over the
  eight validation pairs. Positive values indicate that selected components carry
  more source-conditioned readout effect than matched alternatives.}
  \label{fig:clean_component_specificity}
\end{figure*}
\paragraph{Component-budget sweep.}
We use an auxiliary anger budget sweep to test how quickly the source-to-readout effect appears as
components are added. Retaining only the top 10\% of nodes gives a sufficiency recovery ratio of
0.696 and a necessity removal ratio of 0.661. Increasing the budget to the top 15\% raises these
values to 0.763 and 0.720, and the top 30\% reaches 0.815 and 0.798. Larger budgets
add only modest gains: the top 50\% gives 0.831 recovery and 0.828 removal, while the top 90\%
gives 0.870 recovery and 0.862 removal.

This compression curve supports a sparse component-set interpretation. The effect is not confined to
a single component, but much of the measurable source-to-readout effect is captured by a relatively
small subset of ranked nodes. Group blockout analyses further indicate that late MLPs contribute a
substantial fraction of the effect: removing \(L20\)--\(L23\) MLPs from the anger top-30\% component set
reduces the necessity removal ratio from 0.798 to 0.677 and the sufficiency recovery ratio from
0.815 to 0.698. Removing middle-layer heads also reduces the effect, especially the
head-specific contribution, but the late-MLP removal produces the larger overall decrease in the full
component-set score.

\section{Source-Conditioned Component Specificity Controls}
\label{sec:appendix_component_specificity_controls}

We run an auxiliary component-level control to test whether selected components
outperform cleaner matched alternatives after conditioning-prefix source patching. The layer-0
\([0,32)\) source window remains fixed, and the readout remains the
whole-generation layer-23 residual emotion readout. For each selected head, we
sample five same-layer heads that are not in the frozen selected set. For each
selected MLP, we sample five same-type MLP controls, preferring nearby layers
within \(\pm 2\), expanding to \(\pm 4\), and then filling from the closest
non-selected MLPs if needed. The control manifest records the candidate-pool size,
matching rule, and selected controls for every component.

For each validation pair and selected component, we compare the selected
component score with the mean score of its five controls. We then average these
selected-minus-control deltas over selected components within the same validation
pair and bootstrap over the eight pairs. This keeps the validation pair, rather
than the component row, as the statistical unit. The deltas are positive for all
three emotions and both intervention directions: recovery ranges from \(+0.038\)
to \(+0.064\), and removal ranges from \(+0.038\) to \(+0.074\), with bootstrap
confidence intervals excluding zero in every condition (Figure~\ref{fig:clean_component_specificity}).

\section{Cross-Emotion Component Swaps}
\label{sec:appendix_cross_emotion_robustness}

The canonical component rankings share a common backbone, but the activations at
these sites are emotion dependent. For each ordered source--target emotion pair,
we replace the target-emotion activations at shared sites with donor activations
from the other emotion. Table~\ref{tab:cross_emotion_swaps} reports the target
readout change relative to the target-source intervention. Cross-emotion donors
reduce the target readout in 47 of 48 cases, with an overall mean change of
\(-41.77\) and a 95\% confidence interval excluding zero.

\begin{table}[t]
  \centering
  \scriptsize
  \setlength{\tabcolsep}{3.5pt}
  \caption{\textbf{Cross-emotion shared-component swaps.} Negative values mean
  that another emotion's donor activations weaken the target readout.}
  \label{tab:cross_emotion_swaps}
  \begin{tabular}{lccc}
    \toprule
    \textbf{Target} & \textbf{Mean change} & \textbf{95\% CI} & \textbf{Lower} \\
    \midrule
    Anger & \(-33.14\) & \([-47.44,-18.84]\) & 15/16 \\
    Happiness & \(-40.72\) & \([-52.73,-28.70]\) & 16/16 \\
    Sadness & \(-51.46\) & \([-53.47,-49.46]\) & 16/16 \\
    Overall & \(-41.77\) & \([-48.27,-35.28]\) & 47/48 \\
    \bottomrule
  \end{tabular}
\end{table}

\end{document}